\documentclass[10pt,aps,prd,floatfix,notitlepage,showpacs,nofootinbib]{revtex4-1}

\usepackage{latexsym}
\usepackage{epsfig}
\usepackage{amssymb}
\usepackage{amsmath}
\usepackage{dcolumn}
\usepackage{bm}
\usepackage{multirow}

\newcommand{\lb}{\left[}
\newcommand{\rb}{\right]}

\newcommand{\ba}{\begin{eqnarray}}
\newcommand{\ea}{\end{eqnarray}}
\newcommand{\be}{\begin{equation}}
\newcommand{\ee}{\end{equation}}

\newcommand{\al}{\alpha}

\newcommand{\ga}{\gamma}

\newcommand{\lm}{\mathcal{L}_M}
\begin{document}

\title{On the consistency of non-minimally coupled $f(R)$ gravity}

\author{Nicola Tamanini}\email{n.tamanini.11@ucl.ac.uk}
\affiliation {Department of Mathematics, University College London, Gower Street, London, WC1E 6BT, UK}
\author{Tomi S. Koivisto}\email{tomi.koivisto@fys.uio.no}
\affiliation{Institute of Theoretical Astrophysics, University of Oslo, P.O. Box 1029 Blindern, N-0315 Oslo, Norway}

\date{\today}

\begin{abstract}

Theories with a non-minimal coupling between the space-time curvature and matter fields introduce an extra force due to the
non-conservation of the matter energy momentum. In the present work the theoretical consistency of such couplings is studied using a scalar field
Lagrangian to model the matter content. The conditions that the coupling does not introduce ghosts, classical instabilities or superluminal 
propagation of perturbations are derived. These consistency conditions are then employed to rule out or severely restrict the  
forms of the non-minimal coupling functions considered in the previous literature. For example, a power-law coupling is viable only
for sublinear positive power of the curvature scalar.
\end{abstract}


\maketitle

\section{Introduction}

One of the fundamental issues in gravity is its coupling to the matter fields. In Einstein's general relativity (GR), the equivalence principle dictates the minimal coupling prescription, but despite stringent constraints on its violations at local astrophysical and especially laboratory experiments, there is no compelling reason to take it for granted at all scales and at all times \cite{Will:2005va}. In fact, high energy physics theories often predict nonminimal couplings of matter to e.g.~scalar fields or the gravitational degrees of freedom.

Although the most popular type of non-minimal coupling is the conformal one\footnote{Though more general forms of couplings can be well motivated too, see e.g. \cite{Bekenstein:1992pj,Koivisto:2012za,Brax:2012ie,Bettoni:2013diz}.}, for purely gravitational modifications, it is known that $f(R)$ theories represent the only local, metric-based, generally coordinate invariant and stable modifications of gravity \cite{Woodard:2006nt,Biswas:2013ds}. Thus, when considering the possibility of a non-minimal gravitational coupling, a natural starting point
could be the coupling of a function of the curvature scalar to the matter Lagrangian. The general action for such non-minimally coupled $f(R)$ gravity \cite{Koivisto:2005yk} can be written as
\begin{align}
S_0=\int d^4x\sqrt{-g} \left[f_1(R)+f_2(R)\,\mathcal{L}_M\right] \,,
\label{action:f(R)}
\end{align}
where $f_1(R)$ and $f_2(R)$ are two general (sufficiently regular) functions of the curvature scalar $R$ and $\mathcal{L}_M$ denotes the matter Lagrangian.

Such a theory predicts an extra force due to the non-conservation of the stress energy tensor \cite{Koivisto:2005yk,Bertolami:2007gv}. The resulting non-geodesic motion and some of its astrophysical implications have been extensively analysed in \cite{Bertolami:2007gv,Sotiriou:2008it, Corda:2008uh,Bisabr:2012tg,Bertolami:2008zh,Puetzfeld:2008xu, Mohseni:2010hk,Obukhov:2013ona,Bisabr:2013laa,Puetzfeld:2013ir}. Theories of the form (\ref{action:f(R)}) have been widely applied to cosmology: dark matter has been proposed as an effect of the extra force on baryonic matter \cite{Bertolami:2007gv,Bertolami:2009ic,Harko:2010vs}, dark energy has been considered to arise from the non-minimal coupling \cite{Bertolami:2010cw,Thakur:2010yv}, the evolution of cosmological perturbations has been studied \cite{Nesseris:2008mq,Bertolami:2013kca,Thakur:2013oya}, and inflation was considered in this context too \cite{Bertolami:2010ke}. Astrophysical and Solar System constraints were investigated in \cite{Bertolami:2008im,Bertolami:2007vu,Bertolami:2013qaa,Paramos:2011rw} and spherically symmetric solutions were derived in \cite{Bertolami:2011fz,Bertolami:2013raa}. Connection of the non-minimal coupling with extra dimensional theories was attempted in \cite{Bertolami:2008pa}, and wormholes \cite{MontelongoGarcia:2010xd, Garcia:2010xb,Bertolami:2012fz}, singularities \cite{xu:2012in} and G\"odel-type universes \cite{Zhang:2011zy} have been constructed. The theory (\ref{action:f(R)}) can be considered also in the Palatini formulation \cite{Koivisto:2005yk,Borowiec:2008js,Harko:2010hw}. Furthermore, it was shown to be embedded into the wider framework of C-theories \cite{Amendola:2010bk}, which was then applied to deduce the post-Newtonian parameters for given models \cite{Koivisto:2011tp}. Some other aspects of the viability of the theory and the effective violations of the energy conditions have been already studied in \cite{Bertolami:2009cd,Sotiriou:2008dh,Faraoni:2007sn,Wang:2010bh,Wang:2010zzr}.  Finally, it has been understood that different parameterisations of perfect fluid lagrangians can yield non-equivalent results in the presence of the non-minimal coupling \cite{Bertolami:2008ab,Harko:2008qz,Faraoni:2009rk, Harko:2010zi, Bertolami:2011rb,Minazzoli:2013bva}.

In this paper we study the theoretical consistency of the models (\ref{action:f(R)}). We separate out the propagating degrees of freedom in the non-minimally coupled matter sector, which is intricately mixed with the gravitational degrees of freedom due to the non-minimal coupling. The aim is to analyze whether matter fields may become ghost-like or exhibit unstable growth or superluminality. For this purpose we need to specify the matter field explicitly, and for simplicity we employ a massless scalar field. If the coupling is considered universal, there should be no pathologies for any kind of matter, but in any case the constraints derived from the basic example of a scalar will present necessary conditions for the viability of the coupling function. Moreover, by considering the Lagrangian of a fundamental field instead of employing an effective phenomenological description of an averaged (perfect or otherwise) fluid, our results cannot be compromised by the aforementioned ambiguities of such effective fluid parameterisations.   

We will proceed as follows. In section \ref{sec:actions} we will reformulate the theory in terms of scalar fields and manipulate their action into a convenient form for the consistency analysis. The recipe to uncover the viability of the theory, given the two functions $f_1$ and $f_2$, will then be given in section \ref{sec:consist-conds}. In the following section   \ref{sec:spec_models} we will apply this recipe for several specific cases proposed in the previous literature. Conclusions and discussions will be finally drawn in section \ref{sec:conclusions}. In addition, in the appendix \ref{sec:frlm} we will show that the presented recipe can be applied to theories with general nonlinear dependence upon the matter Lagrangian as well.

{\it Notation and conventions}. In the following we will set $\eta_{\mu\nu}={\rm diag}(-1,+1,+1,+1)$ with the convention $R_{\mu\nu} \sim +\partial_\al \Gamma^\al_{\mu\nu}$, $c=1$ (speed of light) and $M_{\rm Pl}^2=1/(16\pi G)$ (Planck mass).

\section{Equivalent Actions and Conformal Transformations}
\label{sec:actions}

In order to find some constraints on the functions $f_1(R)$ and $f_2(R)$ we will start studying an action which is dynamically equivalent to (\ref{action:f(R)}). The dynamical equivalence of action (\ref{action:f(R)}) with other gravitational theories, expecially scalar-tensor theories, has already been extensively studied \cite{Faraoni:2007sn,Sotiriou:2008it,Bertolami:2008im}. Our approach will closely follow the one exposed in \cite{Bertolami:2008im}. Consider the following theory
\begin{align}
S = \int d^4x\sqrt{-g} \left[f_1(\alpha) +\left(\frac{\partial f_1(\alpha)}{\partial\alpha} +\frac{\partial f_2(\alpha)}{\partial\alpha}\beta\right) \left(R-\alpha\right) +f_2(\alpha)\mathcal{L}_M\right] \,,
\label{action:alfabeta}
\end{align}
where $\alpha$ and $\beta$ are two scalar fields and $f_1$ and $f_2$ are the same functions appearing in (\ref{action:f(R)}). This theory is dynamically equivalent to (\ref{action:f(R)}), i.e.~the classical equations of motion are the same. To prove it we take the variation of (\ref{action:alfabeta}) with respect to $\alpha$ and $\beta$ which yields the following field equations
\begin{align}
\left(\frac{\partial^2f_1}{\partial\alpha^2} +\frac{\partial^2f_2}{\partial\alpha^2}\beta\right) \left(R-\alpha\right) +\frac{\partial f_2}{\partial\alpha} \left(\mathcal{L}_M-\beta\right) &=0 \,,\\
\frac{\partial f_2}{\partial\alpha}\left(R-\alpha\right) &=0 \,.
\end{align}
Provided $f_2\neq{\rm const}$, which anyway represents the minimal coupling, the unique solution of these equations is
\begin{align}
\alpha=R \quad\mbox{and}\quad \beta=\mathcal{L}_M \,.
\label{013}
\end{align}
If we now substitute back into action (\ref{action:alfabeta}) we obtain exactly action (\ref{action:f(R)}) implying that the two actions are indeed dynamically equivalent. In general the second scalar field $\beta$ is unnecessary and we could consider action (\ref{action:alfabeta}) with only one auxiliary field $\alpha$. However the forthcoming analysis is much simplified in the two field approach, which moreover can be straighforwardly applied to more general Lagrangians such as $f(R,\mathcal{L}_M)$ (see the appendix). For these reasons considering two independent auxiliary fields in action (\ref{action:alfabeta}) is more suitable for our purposes.

At this point we define two new scalar fields as
\begin{align}
\phi_1 = \frac{\partial f_1(\alpha)}{\partial\alpha} +\frac{\partial f_2(\alpha)}{\partial\alpha}\beta \quad\mbox{and}\quad \phi_2 = f_2(\alpha) \,,
\label{def:phis}
\end{align}
which allows us to rewrite action (\ref{action:alfabeta}) as
\begin{align}
S=\int d^4x\sqrt{-g} \left[\phi_1R+\phi_2\mathcal{L}_M +V(\phi_1,\phi_2)\right] \,,
\label{action:jordan}
\end{align}
with
\begin{align}
V(\phi_1,\phi_2) = f_1(\alpha)-\phi_1\alpha \,,
\label{009}
\end{align}
where $\alpha$ has to be interpreted as a function depending on $\phi_2$ through (\ref{def:phis}). Action (\ref{action:jordan}) represents a Brans-Dicke theory with vanishing kinetic term and a matter sector coupled to a second scalar field which interacts with the Brans-Dicke field through the potential $V$.

We want now to integrate out the scalar field $\phi_2$. To do this we consider the variation of (\ref{action:jordan}) with respect to $\phi_2$ which yields
\begin{align}
\mathcal{L}_M + \frac{\partial V}{\partial\phi_2}=0 \,.
\label{eq:phi2}
\end{align}
We need a solution of this equation for $\phi_2$. Our analysis only works when Eq.~(\ref{eq:phi2}) admits such a solution. If this is not the case one has to follow other ways in order to constraint the original theory (\ref{action:f(R)}). In any case, as we will see in Sec.~\ref{sec:spec_models}, a solution to (\ref{eq:phi2}) can always be found for a wide class of theories implying that our method is quite generally applicable. Assume then that
\begin{align}
\phi_2^*=\Sigma(\phi_1,\mathcal{L}_M) \,,
\label{001}
\end{align}
is a solution of (\ref{eq:phi2}) with $\Sigma$ a function of $\phi_1$ and $\mathcal{L}_M$. Plugging (\ref{001}) back into action (\ref{action:jordan}) gives
\begin{align}
S=\int d^4x\sqrt{-g} \left[\phi_1R+P(\mathcal{L}_M,\phi_1)\right] \,,
\label{action:P}
\end{align}
where
\begin{align}
P(\mathcal{L}_M,\phi_1)= \mathcal{L}_M\Sigma + V(\phi_1,\Sigma) \,.
\label{008}
\end{align}

We can now perform a conformal transformation in order to switch from the Jordan to the Einstein frame. Consider the following conformally related metric
\begin{align}
\tilde{g}_{\mu\nu} = 16\pi G\,\phi_1\,g_{\mu\nu} \,,
\end{align}
and redefine $\phi_1$ as
\begin{align}
\phi_1 = \phi_0 \exp\left(\frac{\sqrt{G}}{16\pi\sqrt{3}}\tilde\phi_1\right) \,,
\end{align}
where $\phi_0$ is a constant. With the help of this transformation we can rewrite action (\ref{action:P}) as
\begin{align}
S=\int d^4x\sqrt{-\tilde{g}} \left[\frac{R}{16\pi G} -\frac{1}{2}\tilde{g}^{\mu\nu} \partial_\mu\tilde\phi_1\partial_\nu\tilde\phi_1 +\tilde{P}(\mathcal{L}_M,\tilde\phi_1)\right] \,,
\label{action:einstein}
\end{align}
where
\begin{align}
\tilde{P}(\mathcal{L}_M,\tilde\phi_1)= \frac{1}{\phi_1^2} P(\mathcal{L}_M,\phi_1) \,,
\label{006}
\end{align}
with $\phi_1$ to be interpreted as a function of $\tilde\phi_1$ and $\mathcal{L}_M$ depending on $\tilde{g}_{\mu\nu}$ through $g_{\mu\nu}$. Action (\ref{action:einstein}) is nothing but Einstein gravity with a scalar field $\phi_1$ which interacts with all the matter degrees of freedom. This scalar field is obviously regular being its kinetic term of the canonical type. However the matter Lagrangian appears non trivially in the theory and the stability of its fields is not guaranteed.

\section{Consistency Conditions for Non Minimally Coupled $f(R)$ Gravity}
\label{sec:consist-conds}

According to the spirit of the original non minimally coupled $f(R)$ theory given by (\ref{action:f(R)}), one can in principle consider any kind of matter Lagrangian. In other words the theory must be consistent for every choice of matter we want to couple to gravity. In order to continue our analysis we are thus free to choose any matter Lagrangian at our convenience. In what follows we will consider $\mathcal{L}_M$ to be the Lagrangian of a matter scalar field $\phi_M$:
\begin{align}
\mathcal{L}_M \equiv X = -\frac{1}{2}g^{\mu\nu} \partial_\mu\phi_M\partial_\nu\phi_M \,.
\label{matterLag}
\end{align}
In principle we could have added any potential for the scalar field $\phi_M$, but the following analysis would have be not altered and thus we choose to work with only the kinetic term. Note that assuming $\partial_\mu\phi_M$ to be timelike implies that $X>0$. The choice (\ref{matterLag}) is probably the simplest one we can consider and it will allow us to put some constraints on the functions $f_1$ and $f_2$ when a particular model is choosen.
However since we are restricting our study to a specific choice of the matter Lagrangian, the results will be only necessary conditions and more restrictive constraints could be found with different choices. In any case, since the original theory (\ref{action:f(R)}) must work for every matter Lagrangian, any constraint we will obtain on the functions $f_1$ and $f_2$ must be satisfied {\it in primis} by every model. In other words we will be able to exclude the theories which do not satisfy such constraints, but cannot assure that other models are ultimately viable. A complete analysis is impossible for obvious reasons (one should study all the possible matter Lagrangians) and each theory has to be considered separately if one wants to check its consistency in depth.

With the matter Lagrangian choice (\ref{matterLag}) $\tilde{P}$ becomes in general a function of $X$ and $\tilde\phi_1$. The kinetic term for $\tilde\phi_1$ is canonical and separated from $\tilde{P}$, meaning that all the dependence on derivatives of fields inside $\tilde{P}$ is contained in $X$. Such type of theories are known in cosmological contexts as $k$-essence models \cite{ArmendarizPicon:1999rj,Chiba:1999ka,Garriga:1999vw,ArmendarizPicon:2000ah}. They usually depend on a scalar field which enters the action in a non canonical way. The action for $k$-essence is commonly written as
\begin{align}
S_k = \int d^4x\sqrt{-g} \left[\frac{R}{16\pi G} +p(X_\phi,\phi)\right] \,,
\end{align}
where $p$ is a general function of a scalar $\phi$ and $X_\phi=-1/2(\partial\phi)^2$.

Because of their applications to dark energy and inflation, these theories have been well studied and several results have already been discovered. In particular we are interested in the stability conditions one must impose on the function $p$ in order for the model to be physically viable. These can be obtained requiring the positivity of both the energy density of the scalar field and the speed of sound at which the perturbations propagate in the scalar fluid\footnote{Technically this is defined as the ratio of pressure perturbation and the density perturbation evaluated in the rest frame of the field. For a canonical scalar field the speed of sound is identically unity; in general it can depend on time and it differs from the so called adiabatic speed of sound that is the ratio of the time derivative of the background pressure and the background energy density.}. The required conditions are given by \cite{ArmendarizPicon:2000ah,Garriga:1999vw}
\begin{align}
\epsilon(\phi)  =  2X_\phi\,p_{,X_\phi} -p >0 \label{002} \,,\\
c_s^2(\phi)  =  \frac{p_{,X_\phi}}{\epsilon_{,X_\phi}} = \frac{p_{,X_\phi}}{2X_\phi\,p_{,X_\phi X_\phi}+p_{,X_\phi}}  \ge 0 \,,\label{003}
\end{align}
where ${}_{,X_\phi}$ denotes differentiation with respect to $X_\phi$. Here $\epsilon(\phi)$ is the energy density of the scalar field, while $c_s(\phi)$ is the speed of sound. Conditions (\ref{002}) and (\ref{003}) have to be satisfied by every physically viable model. If they do not hold the theory cannot be employed at macroscopic scales.  If we further require that the perturbations do not propagate faster than the speed of light we should impose the further condition that $c_s(\phi) \le 1$.

We now transfer conditions (\ref{002}) and (\ref{003}) to our theory given by action (\ref{action:einstein}). In order for this theory to be consistent we require that
\begin{align}
\tilde\epsilon(\phi_M,\phi_1)  = 2\tilde{X}\,\tilde{P}_{,\tilde{X}}-\tilde{P}>0 \label{004}\,,\\
1 \ge \tilde{c}_s^2(\phi_M,\phi_1) = \frac{\tilde{P}_{,\tilde{X}}} {2\tilde{X}\,\tilde{P}_{,\tilde{X}\tilde{X}}+\tilde{P}_{,\tilde{X}}} \ge 0 \,,\label{005}
\end{align}
where $\tilde{X}$ denotes the kinetic term of the Einstein frame and it is related to $X$ by
\begin{align}
\tilde{X} = \frac{X}{16\pi G\,\phi_1} \,.
\label{007}
\end{align}
The theory can thus be constrained requiring that the conditions (\ref{004}) and (\ref{005}) are satisfied in the Einstein frame. However, since we know the relations between $X$ and $\tilde{X}$ and between $P$ and $\tilde{P}$, we can find how the conditions (\ref{004}) and (\ref{005}) are written in terms of Jordan frame quantities. Using (\ref{006}) and (\ref{007}) we can easily obtain
\begin{align}
\tilde\epsilon = \frac{1}{\phi_1^2}\epsilon \qquad\mbox{and}\qquad \tilde{c}^2_s=c_s^2 \,,
\end{align}
where
\begin{align}
\epsilon = 2X\,P_{,X}-P \qquad\mbox{and}\qquad c_s^2 = \frac{P_{,X}}{2X\,P_{,XX}+P_{,X}} \,,
\label{def:epscs}
\end{align}
are the energy density and speed of sound in the Jordan frame. Thus the two consistency conditions (\ref{004}) and (\ref{005}) can conveniently be translated into
\begin{align}
\epsilon>0 \qquad\mbox{and}\qquad 1 \ge c_s^2 \ge 0 \,,
\label{conds}
\end{align}
which can be computed directly from the Jordan frame avoiding the passage to the Einstein frame. As we will see, once a particular $f(R)$ model is choosen in action (\ref{action:f(R)}), the conditions (\ref{conds}) will permit to constrain the free parameters of the theory.

Before closing this section we conveniently summarize the whole recipe one has to follow in order to find consistency conditions on non minimally coupled $f(R)$ gravity:
\begin{enumerate}
\item Identify the two functions $f_1(R)$ and $f_2(R)$ in action (\ref{action:f(R)});
\item Compute $\phi_1$, $\phi_2$ and $V(\phi_1,\phi_2)$ from definitions (\ref{def:phis}) and (\ref{009});
\item Find a solution $\phi_2^*$ of Eq.~(\ref{eq:phi2}) (if possible);
\item Evaluate $P(\mathcal{L}_M,\phi_1)$ from (\ref{008}) with $\mathcal{L}_M=X$;
\item Constrain the free parameters of the theory using conditions (\ref{conds}).
\end{enumerate}
In the remaining part of the paper we will apply these instructions to some non minimally coupled $f(R)$ models. This will also explain how this procedure works in practice and how it can be employed to constraint the free parameters of the theory.

\section{Consistency of Specific Models}
\label{sec:spec_models}

In this section we will consider some particular models where the functions $f_1(R)$ and $f_2(R)$ of action (\ref{action:f(R)}) are assumed to be of some specific kind. Some of these theories will be considered for their simplicity, other because of their relevance for cosmological and astrophysical applications.

\subsection{$f_1(R)\propto R$ and $f_2(R)\propto R^\gamma$}
\label{model_1}

The first model we study is motivated by its relative simplicity and will be utilized as a working example. We will consider a theory specified by the functions
\begin{align}
f_1(R)=\frac{R}{16\pi G} \qquad\mbox{and}\qquad f_2(R)=A\,R^\gamma \,,
\label{027}
\end{align}
where $A$ and $\gamma$ are two free real parameters. The first function represents nothing but the standard Einstein-Hilbert Lagrangian for GR. Similar to several models already studied in the literature, this specific choice has the scope of only modifying the minimal coupling with matter and to leave the pure gravity sector unchanged. The theory reduces to GR in the limit $\gamma\rightarrow 0$ and $A\rightarrow 1$, and if we want to study only physically motivated modifications of gravity, we cannot consider negative values for $A$ where the gravitational force would be repulsive on all matter fields. In what follows we will thus set $A>0$.
Even though the function $f_2(R)$ is usually taken to be $f_2(R)=1+A\,R^\gamma$ for physical applications of the theory (e.g. \cite{Bertolami:2009ic,Bertolami:2010cw,Bisabr:2012tg,Bertolami:2013kca}), we study first the model (\ref{027}) both because of its simplicity and since it has been already analysed in other works (e.g.~\cite{Wang:2010zzr}). The case $f_2(R)=1+A\,R^\gamma$ will be studied Sec.~\ref{sec:1+powerlaw_model}.

Having identified the two shape functions of the theory, the next step we have to take is to compute the fields $\phi_1$, $\phi_2$ and the potential $V(\phi_1,\phi_2)$. These are given by
\begin{align}
\phi_1= M_{\rm Pl}^2 +\gamma A\, \alpha^{\gamma-1}\beta \,,\qquad \phi_2=A\,\alpha^\gamma \,,\qquad V=\left(M_{\rm Pl}^2-\phi_1\right) \left(\frac{\phi_2}{A}\right)^{1/\gamma} \,,
\label{012}
\end{align}
where $M_{\rm Pl}^2=1/(16\pi G)$ is the Planck mass. At this point we must find a solution to Eq.~(\ref{eq:phi2}) which in this case reads
\begin{align}
\mathcal{L}_M + \left(M_{\rm Pl}^2-\phi_1\right)\frac{1}{\gamma} \frac{1}{A^{1/\gamma}} \phi_2^{(1-\gamma)/\gamma} =0 \,.
\end{align}
Provided $\gamma\neq 1$, which will be analyzed later, we can easily find the solution as
\begin{align}
\phi_2^* = \Sigma(\mathcal{L}_M,\phi_1) = \left(\frac{\gamma A^{1/\gamma} \mathcal{L}_M}{\phi_1-M_{\rm Pl}^2}\right)^{\gamma/(1-\gamma)} \,.
\end{align}
Next we compute $P(\mathcal{L}_M,\phi_1)$ using (\ref{008}). This is given by
\begin{align}
P(\mathcal{L}_M,\phi_1) = (1-\gamma) A^{1/(1-\gamma)} \left(\frac{\gamma}{\phi_1-M_{\rm Pl}^2}\right)^{\gamma/(1-\gamma)} \mathcal{L}_M^{1/(1-\gamma)} \,.
\label{011}
\end{align}
Finally, after setting $\mathcal{L}_M=X$, we are ready to compute the consistency conditions (\ref{conds}) which will constrain the range of possible values of $\gamma$. We first calculate the sound speed which turns out as
\begin{align}
c_s^2 = \frac{1-\gamma}{1+\gamma}\,.
\label{010}
\end{align}
For absence of instabilities we thus require that
\begin{align}
c_s^2 \ge 0 \quad\Leftrightarrow\quad -1 \le \gamma \le 1 \,.
\label{017}
\end{align}
This tells us that values of $\gamma$ for which $|\gamma|\geq 1$ are forbidden in such a theory. This is expected since small deviations from GR, for which $\gamma=0$, are usually more viable under a phenomenological point of view. If we further wish to avoid superluminal signals, we obtain the constraint
\begin{align}
0 \le c_s^2 \le 1 \quad\Leftrightarrow\quad 0 \le \gamma \le 1 \,.
\label{signal}
\end{align}
The constraint coming from the positivity of the energy density gives us
\begin{align} \label{Pcon}
\epsilon = \frac{1+\gamma}{1-\gamma}\,P>0 \,,
\end{align}
which combined with (\ref{010}) leaves us with
\begin{align}
P>0 \,.
\end{align}
Since $A>0$, $\mathcal{L}_M=X>0$ and $|\gamma|<1$, the only object which remain to check in (\ref{011}) is $\gamma/(\phi_1-M_{\rm Pl}^2)$. From the definition of $\phi_1$ (\ref{012}) we have
\begin{align}
\frac{\gamma}{\phi_1-M_{\rm Pl}^2} = \frac{1}{A\,\alpha^{\gamma-1}\beta} \,,
\end{align}
whose left hand side is determined by the right hand side which, in the GR limit $\gamma\rightarrow 0$, $A\rightarrow 1$ and $\phi_1\rightarrow M_{\rm Pl}^2$, becomes
\begin{align}
\left.\frac{\gamma}{\phi_1-M_{\rm Pl}^2}\right|_{GR} \rightarrow \frac{\alpha}{\beta} \approx \frac{R}{\mathcal{L}_M} \,,
\end{align}
where in the last equality the solution (\ref{013}) of the equations of motion has been used. We can then realize that the fraction $\gamma\rightarrow 0$ over $\phi_1- M_{\rm Pl}^2\rightarrow 0$ controls the relative sign between $R$ and $\mathcal{L}_M$. Since in GR this is nothing but the Planck mass $M_{\rm Pl}$, it is natural to assume that also for physically viable theory this remains positive. In this manner there are no inconsistency in the definition of $P$ (\ref{011}) and the positive energy condition is automatically satisfied.

This model is thus always unphysical whenever $|\gamma|>1$ and presents superluminal propagation unless $0<\gamma<1$. Employing our recipe we managed to impose a constraint on the range of the possible values of the free parameter $\gamma$.

\subsubsection{Remarks on the linear coupling and viable (and otherwise) generalizations of the theory}
\label{remarks}

It is interesting to stop to consider briefly the special case of the linear coupling $\ga=1$ because of its particular simplicity (phenomenological studies have also claimed the linear coupling may be observationally viable in some regimes \cite{Bertolami:2007vu}) and because it has been excluded from the analysis above. In this particular case, it is possible to make easily contact with the Horndeski theories and thus propose viable generalizations of the theory. 

Here the ghost condition (\ref{Pcon}) appears to become ill-defined
at the limit of linear coupling. However, by comparing with the Horndeski theories we know that a term $\sim R(\partial\phi)^2$ in the action entails a ghost\footnote{Such couplings were studied already by Amendola in \cite{Amendola:1993uh}. The more recent nonminimally coupled three-form model \cite{Koivisto:2012xm} might also have ghost, as in the scalar field dual description the three-form field invariant is the kinetic term of the scalar.}. Thus $\gamma=1$ is ruled out. If one instead considers the full Horndeski term $\sim R(\partial\phi)^2-2[(\Box\phi)^2-\phi_{;\mu\nu}\phi^{;\mu\nu}]$, the higher derivatives in the equations of motion are canceled and the energy associated with the fluctuations of the field can remain bounded from below despite the appearance of the higher derivatives in the action. Thus, modifying the coupling in this way would cure the negative-energy pathology. 

Written in terms of the perfect fluid parameterisation, where one identifies the properly normalized four-velocity as $u_\mu=\phi_{,\mu}/\sqrt{2X}$, the coupling would look rather clumsy,
\be
R\lm +  \frac{1}{2}\lm \lb (u\cdot\nabla\log{\lm})^2+(\nabla\log{\lm})^2\rb + 2 \lm \lb (\nabla\cdot u)(u\cdot\log{\lm})+(\nabla\cdot u)^2-(\nabla u)^2\rb\,.
\ee
A lesson here is at least that when writing down nonminimally coupled theories, especially with extra derivatives (either the $\nabla$'s acting directly on the matter lagrangian or 
the derivatives implicit in the curvature scalar), one has to consider very well specified classes of theories in order to avoid the introduction of unphysical pathologies. This is extremely useful since it permits to efficiently constrain the otherwise infinite theory space and to immediately exclude models such as e.g.~those considered in\footnote{Some of these and many other studies start with an action involving the stress energy tensor (contracted with the metric or some other tensor). It is difficult to see how this could make sense conceptually, as the stress energy tensor is fundamentally {\it derived} from the action. However, it is possible to consider theories constructed through recursive definitions for the couplings of the stress energy tensor \cite{Singh:1988pp,Sami:2002se}, thus entailing a similar loopy structure that emerges in C-theories \cite{Amendola:2010bk} and nonlocal gravity \cite{Sandstad:2013oja}.} \cite{Poplawski:2006ey,Alvarenga:2013syu, Harko:2012hm,Odintsov:2013iba}.

\subsection{$f_1(R)\propto R$ and $f_2(R)=1+ (R/R_0)^\gamma$}
\label{sec:1+powerlaw_model}

In this section we study a model similar to the one already analyzed in Sec.~\ref{model_1}. The non-minimal coupling is again characterized by a power law function, but now the matter Lagrangian appears also with its canonical terms:
\begin{align}
f_1(R)=\frac{R}{16\pi G} \qquad\mbox{and}\qquad f_2(R)=1+\left(\frac{R}{R_0}\right)^\gamma \,,
\end{align}
where $R_0$ and $\gamma$ are here the parameters of the model. The GR limit is given by $R/R_0\ll 1$ but also the limit $\gamma\rightarrow 0$ reduces the theory to standard GR after a redefinition of the gravitational constant.
Though this model is mathematically almost the same of the one in Sec.~\ref{model_1}, physically it presents more interesting features because the deviations from standard GR can be better parametrized. For this reason it has been studied more intensively through the literature \cite{Bertolami:2009ic,Bertolami:2010cw,Bisabr:2012tg,Bertolami:2013kca}.

The scalar fields (\ref{def:phis}) and the potential (\ref{009}) are now
\begin{align}
\phi_1= M_{\rm Pl}^2 +\frac{\gamma}{R_0^\gamma} \, \alpha^{\gamma-1}\beta \,,\qquad \phi_2=1+\left(\frac{\alpha}{R_0}\right)^\gamma \,,\qquad V=R_0\left(M_{\rm Pl}^2-\phi_1\right) \left(\phi_2-1\right)^{1/\gamma} \,,
\label{015}
\end{align}
and solving Eq.~(\ref{eq:phi2}) eventually yields
\begin{align}
P(X,\phi_1) = X+(1-\gamma)\,X^{1/(1-\gamma)}\left(\frac{\gamma}{R_0\,(\phi_1-M_{\rm Pl}^2)}\right)^{\gamma/(1-\gamma)} \,.
\end{align}
As expected, this is, a part different definition of the parameters, the same of (\ref{011}) but with a canonical term added.

Before evaluating conditions (\ref{conds}) we give an argument for the positiveness of
\begin{align}
F_\gamma(\phi_1) = \left[\frac{\gamma}{R_0\,(\phi_1-M_{\rm Pl}^2)}\right]^{\gamma/(1-\gamma)} \,.
\end{align}
If we require $R_0>0$ for consistency then from the definition of $\phi_1$ (\ref{015}) we obtain
\begin{align}
\frac{\phi_1-M_{\rm Pl}^2}{\gamma} = \frac{\beta}{R_0}\left(\frac{\alpha}{R_0}\right)^{\gamma-1} \approx \frac{\mathcal{L}_M}{R_0}\left(\frac{R}{R_0}\right)^{\gamma-1} \,,
\label{016}
\end{align}
where in the last equality the solution (\ref{013}) of the equations of motion has been used. In this case the GR limit $R/R_0\ll 1$ is not very insightful since just forces $\phi_1\approx M_{\rm Pl}^2$ as it has to be. However, as we noticed above, the limit $\gamma\rightarrow 0$ also reduces the theory to GR. In this limit the left hand side of (\ref{016}) is indeterminate while the right hand side yields again $\mathcal{L}_M/R$. This implies that in the GR limit the left hand side is nothing but the relative sign between $R$ and $\mathcal{L}_M$, or in other words, the gravitational constant. Since in GR this constant is positive it is again natural to assume that physically motivated deviations from GR also leaves this constant positive. In what follows we can thus assume that $F_\gamma(\phi_1)>0$.

We can now compute the energy and speed of sound from (\ref{def:epscs}) obtaining
\begin{align}
\epsilon = X+(1+\gamma)\,F_\gamma(\phi_1)\,X^{1/(1-\gamma)} \,,\\
c_s^2 = \frac{1+F_\gamma(\phi_1)\,X^{\gamma/(1-\gamma)}} {1+\frac{1+\gamma}{1-\gamma}F_\gamma(\phi_1)\,X^{\gamma/(1-\gamma)}} \,.
\end{align}
In general it is now more complicate to understand for which values of $\gamma$ the conditions (\ref{conds}) are not satisfied. However such conditions have to be true for every value of the kinetic energy $X$. Requiring that the conditions (\ref{conds}) are satisfied both when $X\gg 1$ and $X\ll 1$ provide the following constraint
\begin{align}
-1 \le \gamma \le 1 \,,
\end{align}
which must be true in order for the theory to be viable. Furthermore, excluding superluminal propagation cuts out the possibility of a negative exponent and the constraint
tightens to $0 \le \gamma < 1$. 
Note that this is the same result we have obtained in (\ref{017}). This does not come as a surprise since in one of the regimes $X\gg 1$ or $X\ll 1$ the canonical kinetic term of the scalar field is always negligible in comparison with the non-minimal coupling term. The theory effectively becomes the same as the one analysed in Sec.~\ref{model_1}.

Since the inclusion of the canonical matter term inside the gravitational action just corresponds to the redefinition $P(X,\phi_1)\mapsto X+P(X,\phi_1)$ in our analysis, we can conjecture that
every conditions valid for a theory $f_2(R)=f^*(R)$ are also true for a theory $f_2(R)=1+f^*(R)$ with $f^*(R)$ a general function. This is because in the appropriate limit of the kinetic energy of the matter scalar field the canonical matter term will always result negligible if compared to the non-minimal coupled term.

\subsection{$f_1(R)\propto R$ and $f_2(R)\propto e^{\lambda R}$}
\label{sec:exp_model}

Another simple model to analyse is the exponential non-minimal coupling described by
\begin{align}
f_1(R)=\frac{R}{16\pi G} \qquad\mbox{and}\qquad f_2(R)=A\,e^{\lambda R} \,,
\label{014}
\end{align}
where $\lambda$ and $A$ are two free parameters. The GR limit is now characterized by $\lambda\rightarrow 0$ and $A\rightarrow 1$ which suggests to consider $A>0$ for consistency. The two scalar fields (\ref{def:phis}) and the potential (\ref{009}) are now given by
\begin{align}
\phi_1= M_{\rm Pl}^2 +\lambda A\, e^{\lambda\alpha}\beta \,,\qquad \phi_2=A\,e^{\lambda\alpha} \,,\qquad V=\frac{M_{\rm Pl}^2-\phi_1}{\lambda} \log\left(\frac{\phi_2}{A}\right) \,,
\end{align}
and Eq.~(\ref{eq:phi2}) reads
\begin{align}
\mathcal{L}_M + (M_{\rm Pl}^2-\phi_1)\frac{1}{\lambda\,\phi_2} = 0 \,,
\end{align}
with solution
\begin{align}
\phi_2^* = \frac{M_{\rm Pl}^2-\phi_1}{\lambda\,\mathcal{L}_M} \,.
\end{align}
At this point one can easily compute $P(X,\phi_1)$ to be
\begin{align}
P(X,\phi_1) = \frac{M_{\rm Pl}^2-\phi_1}{\lambda} \left(1-\log\left[\frac{A\,\lambda\,X}{M_{\rm Pl}^2-\phi_1}\right]\right) \,.
\end{align}
Finally it is enough to evaluate the condition on the sound speed in order to realize that this model is immediately ruled out. In fact from (\ref{def:epscs}) we find
\begin{align}
c_s^2 = -1 \,,
\end{align}
which is obviously never positive.

To conclude we have found that the non-minimal coupled $f(R)$ model characterized by (\ref{014}) is never physically viable. It is thus not possible to employ such a theory in order to study consistent modifications of gravity at large scales. Of course the case $\lambda=0$ corresponding to GR does not present any problem and is indeed excluded from this analysis.
Finally we mention that also the model $f_2(R)=1+A\exp(\lambda R)$ is ruled out by our analysis. In fact in the appropriate limit of kinetic energy this theory would become identical to the one we just ruled out. Since we require that a non-minimally coupled theory is physically viable for every value of the matter Lagrangian one considers, adding a canonical matter term to such a model will not prevent these instabilities.

\subsection{$f_1(R)=\frac{R}{16\pi G}+A\,R^n$ and $f_2(R)\propto R^n$}
\label{sec:non_EH}

Finally an example where the function $f_1$ is different from the usual Einstein-Hilbert term will be provided in this section. We will generalize the powerlaw model of Sec.~\ref{model_1} by adding a similar self-interacting powerlaw term to the gravitational sector. The two functions in (\ref{action:f(R)}) will then be choosen to be
\begin{align}
f_1(R)=\frac{R}{16\pi G}+A\,R^n \quad\mbox{and}\quad f_2(R)=B\,R^n \,,
\label{021}
\end{align}
where $A$, $B$ and $n$ are three parameters. In general, in accordance with other works \cite{Bertolami:2009cd}, we could have considered different exponents for the powerlaw terms in the two functions. However setting them to coincide enormously simplify the analysis of this section and since our scope here is to present a simple example where the gravitational Lagrangian differs from Einstein-Hilbert, we will only consider the functions (\ref{021}) in the following. Note that the GR limit is achived for $A\rightarrow 0$, $B\rightarrow 1$ and $n\rightarrow 0$.

In the present case the two scalar fields defined in (\ref{def:phis}) can be computed to be
\begin{align}
\phi_1=M_{\rm Pl}^2+n\alpha^{n-1}(A+B\,\beta) \quad\mbox{and}\quad \phi_2=B\,\alpha^n \,,
\label{024}
\end{align}
while the potential (\ref{009}) is given by
\begin{align}
V(\phi_1,\phi_2)= (M_{\rm Pl}^2-\phi_1)\left(\frac{\phi_2}{B}\right)^{1/n} +\frac{A}{B}\phi_2 \,.
\end{align}
At this point we can find Eq.~(\ref{eq:phi2}) as
\begin{align}
\mathcal{L}_M+\frac{1}{n}B^{-1/n}\,(M_{\rm Pl}^2-\phi_1)\,\phi_2^{(1-n)/n} +\frac{A}{B} =0 \,,
\label{022}
\end{align}
with solution
\begin{align}
\phi_2^*=\Sigma(\mathcal{L}_M,\phi_1) = B\left[\frac{n\,B\left(\mathcal{L}_M+A/B\right)} {\phi_1-M_{\rm Pl}^2}\right]^{n/(1-n)} \,.
\end{align}
Eq.~(\ref{022}) would have been much more difficoult to solve analytically if the two exponents in the functions (\ref{021}) had been different.
One can now find $P(X,\phi_1)$ to be
\begin{align}
P(X,\phi_1)= B\,(1-n)\left(\frac{n\,B}{\phi_1-M_{\rm Pl}}\right)^{n/(1-n)} \left(X+\frac{A}{B}\right)^{1/(1-n)} \,.
\end{align}
The speed of sound will then results
\begin{align}
\frac{1}{c_s^2} = 1+\frac{2n}{(1-n)}\frac{X}{(X+A/B)} \,.
\end{align}
Since this must be positive for every value of $X$, when $X\gg 1$ we obtain the condition
\begin{align}
-1 \le n \le 1 \,,
\label{023}
\end{align}
which is the same we obtained in Sec.~\ref{model_1}. Again also the superluminality constraint excludes negative exponents and we are left with $0\le n \le 1$.
On the other hand the positivity of the energy density become
\begin{align}
\epsilon = B\,(1-n)\left(\frac{n\,B}{\phi_1-M_{\rm Pl}}\right)^{n/(1-n)} \left(X+\frac{A}{B}\right)^{n/(1-n)} \left(X\frac{1+n}{1-n}-\frac{A}{B}\right) >0 \,.
\label{026}
\end{align}
To check the positivity of this expression we need to discuss all the terms appearing in it. Due to (\ref{023}) the term $(1-n)$ is always positive. Moreover we can assume that also the third term is positive because of the following argument. From the definitions (\ref{024}) we have
\begin{align}
\frac{\phi_1-M_{\rm Pl}^2}{n\,B}=\left(\beta+\frac{A}{B}\right)\alpha^{n-1}\,,
\end{align}
which in the GR limit reduces to
\begin{align}
\left.\frac{\phi_1-M_{\rm Pl}^2}{n\,B}\right|_{GR}= \frac{\beta}{\alpha}\approx \frac{\mathcal{L}_M}{R}\,,
\end{align}
where in the last equality the solution (\ref{013}) has been used.
Again, since in GR this reduces to the relative sign between the gravitational and matter sector, which is basically nothing but the gravitational constant, it is natural to assume that also for physically justified modifications of gravity this value remains positive. Moreover we notice that
\begin{align}
\frac{\phi_1-M_{\rm Pl}^2}{n\,B}\left(\beta+\frac{A}{B}\right)^{-1} =\alpha^{n-1}\approx R^{n-1}\,.
\label{025}
\end{align}
The RHS of this equation is always positive because the theory we are dealing with is well-defined only for values $R>0$. If $R<0$ the two functions (\ref{021}) are ill-defined and a different definition should be employed. In any case the RHS of (\ref{025}) would always remain positive and thus we can safely assume that also the fourth term in (\ref{026}) is bigger than zero.

The only terms that remain to check in (\ref{026}) are the first and the last one. If $A$ and $B$ have different sign the last term is always positive, whilst if $A/B>0$ it can be positive or negative depending on the value of $X$. Since we require the energy to be positive for all values of $X$, we are led to consider only the case $A/B<0$. However the first term in (\ref{026}) is just $B$ and we must assume $B>0$ if we want $\epsilon>0$. This in turn implies $A<0$. Note that the assumption $B>0$ is quite natural if we require the theory to be physically interesting since in the GR limit we must have $B=1$.

Finally we comment on the fact that we could have choosen $f_2(R)=1+B\,R^n$. As before this redefinition just complicates the whole analysis but does not change the final results. Note that in this case the requirement $B>0$ has nothing to do with the GR limit of the theory because this is $B\rightarrow 0$ now. In this case the $B>0$ condition is indeed a constraint arising from the result of our analysis and has nothing to do with the physical features of the theory.

\section{Discussion and Conclusion}
\label{sec:conclusions}

The purpose of the present work has been to constrain non-minimally coupled $f(R)$ gravity. In order to achieve the goal, the dynamically equivalent representation of the theory in terms of multi-scalar-tensor theories has been employed. By choosing the matter Lagrangian to be a simple scalar field, it has been possible to compute the effective energy density and the speed of sound of adiabiatic perturbations. Finally, to ensure that the non-minimal couplings of the theory do not introduce instabilities it was eventually required that these quantities be positive, and to exclude superluminal propagation of perturbations the sound speed was bounded below unity. The whole procedure has given rise to a recipe, summarized at the end of Sec.~\ref{sec:consist-conds}, which can be adopted to constrain possible models of non-minimally coupled $f(R)$ gravity.
In order to better explain how this recipe works, some specific models have been studied in Sec.~\ref{sec:spec_models}. The main results of this analysis are schematized in Table \ref{tab}.
\begin{table}[ht]
\begin{center}
\begin{tabular}{|c|c|c|c|}
\hline
$f_1(R)$            & $f_2(R)$              & Stable if & Stable \& causal if  \\
\hline
\hline
\multirow{2}*{$\frac{R}{16\pi G}$} & $A\,R^\gamma$  & \multirow{2}{*}{$-1\le \gamma<1$}  &  \multirow{2}{*}{$0 \le \gamma<1$}  \\
 & $1+A\,R^\gamma$  & & \\ 
\hline
\multirow{2}*{$\frac{R}{16\pi G}$} & $A\,\exp(\lambda\,R)$  & \multirow{2}*{$\lambda=0$} &  \multirow{2}*{$\lambda=0$} \\
 & $1+A\,\exp(\lambda\,R)$  & & \\
\hline
\multirow{2}*{$\frac{R}{16\pi G}+A\,R^n$} & $B\,R^n$ & $-1 \le n<1$ & $0 \le n<1$ \\
        & $1+B\,R^n$ & $A<0$ and $B>0$  & $A<0$ and $B>0$ \\
\hline
\end{tabular}
\end{center}
\caption{Constraints on parameters of some specific models of action (\ref{action:f(R)}) arising from the analysis presented in Sec.~\ref{sec:spec_models}.}
\label{tab}
\end{table}
As it is clear from the table, the parameters of the various models have been constrained to lie in specific ranges.

In Sec.~\ref{model_1} and Sec.~\ref{sec:1+powerlaw_model} a powerlaw non-minimal coupling has been considered. This kind of model is particularly popular in literature \cite{Bertolami:2009ic,Bertolami:2010cw,Bisabr:2012tg,Bertolami:2013kca}, both because it is sufficiently easy to handle and since it produces results comparable with observations.
The results of Sec.~\ref{sec:1+powerlaw_model} are consistent with \cite{Sotiriou:2008dh} which poses doubts on the viability of a linear Ricci coupling to matter, though for astrophysical applications such a coupling can be preferable \cite{Bertolami:2007vu}. We argued that the linear coupling is excluded, and proposed a viable generalisation of the theory. In any case, for dark matter and dark energy modeling is commonly considered $\gamma<0$ \cite{Bertolami:2009ic,Bertolami:2010cw}, since one wants the effects of the non-minimal coupling to be relevant at late times when the Ricci scalar is small. However, such models allow superluminal propagation and thus, apparently, violations of causality. The galactic rotation curves have been explained employing $\gamma=-1/3$ or $\gamma=-1$ \cite{Bertolami:2009ic}, while for dark energy values $\gamma<-1$ better fit the observations \cite{Bertolami:2010cw}. Since from the analysis of Sec.~\ref{sec:1+powerlaw_model} it emerges a constraint $\gamma \geq 0$, it seems that a powerlaw non-minimally coupled $f(R)$ model can never represent a suitable late time theoretical description for dark matter or dark energy.

In Sec.~\ref{sec:exp_model} an exponential non-minimal coupling has been probed. Despite its simplicity it has been shown to always lead to a negative speed of sound for adiabatic perturbations. The model has thus been ruled out by the considerations presented in this work and should not be considered for physical applications. As it is shown in Table \ref{tab}, the only possible value for the parameter $\lambda$ is zero which corresponds to nothing but GR. Finally, in Sec.~\ref{sec:non_EH}, a model where also the gravitational sector is modified has been considered. The results for this case are similar to the ones obtained in Sec.~\ref{model_1} with the addition that the parameters $A$ and $B$ have always to be negative and positive, respectively.

To conclude, the analysis exposed in these pages aims at presenting and explaining a simple procedure one can utilize in order to find constraints on specific models of non-minimally coupled $f(R)$ gravity. Though for complicated models this can be practically impossible to carry out, for sufficiently simple cases, such as the ones commonly studied in literature and considered in Sec.~\ref{sec:spec_models}, it becomes a useful tool to check phenomenological viability and rule out unphysical theories. Having in mind the possibility of employing such theories in astrophysical and cosmological applications, it is convenient to have an instrument able to determine the consistency of the model one chooses to work with.


\appendix

\section{The Scalar-Tensor Representation of $f(R,\mathcal{L}_M)$ Gravity}
\label{sec:frlm}

In this appendix we will show that an equivalent scalar-tensor representation exists also for more general theories\footnote{For remarks on other generalizations, recall section \ref{remarks}.} identified by the extended action
\begin{align}
S_0 = \int d^4x\sqrt{-g}\, f(R,\mathcal{L}_M) \,,
\label{018}
\end{align}
where now $f$ is an arbitrary function of both the Ricci scalar $R$ and the matter Lagrangian $\mathcal{L}_M$.
Such models have been proposed in \cite{Harko:2010mv} as a maximal extension of nonminimally coupled $f(R)$ gravity and have been further studied in \cite{Harko:2012ve,Wang:2012tm,Huang:2013dca}. In what follows action (\ref{018}) will be proven to be dynamically equivalent to action (\ref{action:jordan}) with different definitions of the scalar fields $\phi_1$, $\phi_2$ and the potential $V(\phi_1,\phi_2)$. Once this equivalence has been defined, the procedure following (\ref{action:jordan}) can be identically repeated and the general conclusions we found for non-minimally coupled $f(R)$ gravity will also hold for $f(R,\mathcal{L}_M)$ gravity. The only crucial difference will be in the definitions of the scalar fields which in turns will determine the physical differences between the various models.

Consider again another action given by
\begin{align}
S = \int d^4x \sqrt{-g} \left[f(\alpha,\beta)+\frac{\partial f}{\partial\alpha}\left(R-\alpha\right)+\frac{\partial f}{\partial\beta}\left(\mathcal{L}_M-\beta\right)\right] \,,
\label{019}
\end{align}
where $\alpha$ and $\beta$ are two scalar fields. The variation of this action with respect to the two scalar fields gives respectively
\begin{align}
\frac{\partial^2f}{\partial\alpha^2}\left(R-\alpha\right) +\frac{\partial^2f}{\partial\alpha\partial\beta} \left(\mathcal{L}_M-\beta\right) = 0 \,,\\
\frac{\partial^2f}{\partial\alpha\partial\beta}\left(R-\alpha\right) +\frac{\partial^2f}{\partial\beta^2} \left(\mathcal{L}_M-\beta\right) = 0 \,.
\end{align}
Given that the determinant of this system of equations is non vanishing, i.e.~assuming that
\begin{align}
\frac{\partial^2f}{\partial\alpha^2} \frac{\partial^2f}{\partial\beta^2} \neq \left(\frac{\partial^2f}{\partial\alpha\partial\beta}\right)^2 \,,
\end{align}
the unique solution is
\begin{align}
\alpha = R \quad\mbox{and}\quad \beta=\mathcal{L}_M \,.
\end{align}
It is now easy to realize that substituting back this solution into (\ref{019}) immediately produce action (\ref{018}), implying that the two theories are dynamically equivalent.

At this point we define two new scalar fields and a potential as
\begin{align}
\phi_1 = \frac{\partial f}{\partial\alpha} \,,\qquad \phi_2=\frac{\partial f}{\partial\beta} \,,\qquad V(\phi_1,\phi_2)=f(\alpha,\beta)-\alpha\phi_1-\beta\phi_2 \,,
\label{020}
\end{align}
where one must consider $\alpha$ and $\beta$ as functions of $\phi_1$ and $\phi_2$ in the potential $V$. In terms of these new variables action (\ref{019}) can be rewritten as
\begin{align}
S=\int d^4x\sqrt{-g} \left[\phi_1R+\phi_2\mathcal{L}_M +V(\phi_1,\phi_2)\right] \,,
\end{align}
which coincides with (\ref{action:jordan}) but for the definitions (\ref{020}) which differs from (\ref{def:phis}) and (\ref{009}).

To conclude we have proven that the extended theory defined by action (\ref{018}) can equally be mapped into a scalar-tensor theory described by action (\ref{action:jordan}). The only differences arise in the definitions of the scalar fields $\phi_1$, $\phi_2$ and the potential $V$ which generalize the corresponding definitions in non-minimally coupled $f(R)$ gravity. Because of this result the same analysis we performed in the main body of the paper could be identically carried out for $f(R,\mathcal{L}_M)$ gravity.

\begin{acknowledgments}
TK is supported by the Norwegian Research Council of Norway.
\end{acknowledgments}
\appendix

\end{document}